\documentclass[aps,pra,reprint, amsmath, amssymb,superscriptaddress]{revtex4-1}

\usepackage{bm}
\usepackage[retainorgcmds]{IEEEtrantools}
\usepackage{graphicx}
\usepackage{mathrsfs}
\usepackage{amsmath}
\usepackage{amsfonts}
\usepackage{amssymb}
\usepackage{color}
\usepackage{times,txfonts}
\usepackage{nicefrac}
\usepackage{physics}
\usepackage{ragged2e}
\usepackage{tikz}
\usepackage{bbold}
\usepackage[colorlinks=true,linkcolor=blue,urlcolor=blue,citecolor=blue,pdfusetitle]{hyperref}
\usetikzlibrary{arrows,snakes,backgrounds,patterns}

\definecolor{black}{rgb}{0.368417, 0.506779, 0.709798}
\definecolor{Grey}{rgb}{0.43, 0.5, 0.5}
\definecolor{Red}{rgb}{0.922526, 0.385626, 0.209179}
\definecolor{Yellow}{rgb}{1.0, 0.75, 0.0}
\definecolor{Green}{rgb}{0.37820393249936934, 0.6, 0.6}
\definecolor{Purple}{rgb}{0.6171875, 0.26171875, 0.58203125}
\definecolor{purp}{rgb}{0.54, 0.17, 0.89}

\newcommand{\Var}{\text{var}}

\usepackage{soul}
\usepackage{xcolor}
\usepackage[authormarkup=none]{changes}
\definechangesauthor[name={Gabriel}, color=red]{g}
\definechangesauthor[name={Marcelo}, color=purp]{m}

\begin{document}

\title{Quantum mean-square predictors and thermodynamics}
\date{\today}
\author{Marcelo Janovitch}
\email{marcelobro.pera@gmail.com}
\affiliation{Department of Physics, University of Basel, Klingelbergstrasse 82, 4056 Basel, Switzerland.}
\affiliation{Instituto de F\'isica da Universidade de S\~ao Paulo,  05314-970 S\~ao Paulo, Brazil.}
\author{Gabriel T. Landi}
\email{gtlandi@gmail.com}
\affiliation{Instituto de F\'isica da Universidade de S\~ao Paulo,  05314-970 S\~ao Paulo, Brazil.}

\begin{abstract}
Thermodynamic quantities, such as heat and work, are not functions of state, but rather of the process undergone by a physical system. 
Assessing them can therefore be difficult, since it requires  probing the system  at least twice.
This is even more so when these quantities are to be assessed at the stochastic level. 
In this letter we show how to obtain optimal estimates of thermodynamic quantities solely from indirect measurement unravellings of an auxiliary system.
The method always yields the true average. 
And the mean-squared error of the prediction is directly proportional to how well the method estimates the variance.
As an application, we study energy fluctuations in a driven system, and in an avoided crossing work protocol.

\end{abstract}
\maketitle{}

\section{Introduction} 

At the nanoscale, thermodynamic quantities, such as heat and work, may fluctuate significantly~\cite{Evans1993,*Evans1994,Gallavotti1995b,Jarzynski1997,Jarzynski1997a,Crooks1998,Kurchan1998,Jarzynski2004a,Talkner2007,Esposito2009,Campisi2011}.
Properly accounting for these fluctuations is crucial, both from a fundamental as well as from an applied perspective. 
For instance, in nanoscale engines both the output power~\cite{Pietzonka2017,Denzler2020} and the efficiency~\cite{Verley2014a,Denzler2019,Denzler2021} may fluctuate significantly, and the consequences of this are only now starting to be explored.

A unique feature of thermodynamics, however, is that said quantities do not depend on the state of the system, but rather on the process/transformation in question.
Within a quantum setting, several approaches have been developed for tackling this problem~\cite{Allahverdyan2005a,Engel2007,Allahverdyan2014,Hofer2016,Solinas2016,Hofer2017,Miller2017,Solinas2017,YungerHalpern2018a,Levy2019,Park2017,Micadei2020,Guerardini2020}.
But still today, the most widely used is the  two-point measurement  (TPM) scheme~\cite{Talkner2007,Esposito2009,Campisi2011}, where a projective measurement in the system energy basis is applied before and after the process.
Very often, however, the ``system'' is actually composed of multiple parts. 
For instance, in the case of a heat engine operating between two baths, the full statistics of heat and work is in general only accessible by performing a TPM in both the working fluid \emph{and} the two baths~\cite{Talkner2009}, which can be  prohibitive.

When not all parts of the system are accessible, it becomes necessary to develop strategies to estimate thermodynamic quantities indirectly~\cite{Miller2017,Mohammad2018,Manzano2018a,Deffner2016,Sone2018}.
A concrete experimental example is the  calorimetric method developed in~\cite{Pekola2013,Suomela2016,Karimi2020}, which estimates work in a quantum system by measuring the heat flowing to an ancilla (which acts as a finite reservoir).
This paradigm is also quite frequent in open quantum system, e.g. in the study of Full Counting Statistics~\cite{Esposito2009,Levitov1993,Agarwalla2018,Saryal2020,Guarnieri2017a}, where the heat statistics is determined from measurements in the bath.

Motivated by this issue, in this letter we approach the problem from the angle of statistical inference.
We consider a system $S$ undergoing a generic open process in contact with an ancillary system $A$, which can have any size.
It is assumed that the system can never be measured; however, one has access to a certain set of outcomes of the ancilla, corresponding to the unravellings of the open system dynamics~\cite{Carmichael2015, Wiseman2009}.
Our goal is to estimate the \emph{changes} (at the stochastic level) in some system observable $G_t$, between two different times (e.g. $0$ and $\tau$).
More concretely, we ask \emph{what is the best possible prediction one can make about the changes in $G$, given only stochastic outcomes in the ancilla?}
We formulate our results  using the notion of statistical predictors~\cite{Bickel1977}.
Being inference-based, our method is thus directly applicable to experiments.
Our main result is Eq.~\eqref{pred_opt}, which specifies the optimal mean-squared predictor as function only of the Kraus operators determining the process. 
This, as we show, can then be directly applied to thermodynamic protocols, e.g. for the estimation of work. 
However, the result also holds for  any system operator, and thus  extends beyond thermodynamics.
To illustrate the ideas, we consider the estimation of energy fluctuations in a driven qubit, and the determination of work in an avoided crossing protocol.


\section{Statement of the problem }
We consider a system $S$ prepared in a state $\rho_S$, interacting with an ancillary system $A$, prepared in a state  $\rho_A$. 
The total Hamiltonian is taken to have the general form $H(t) = H_S(t) + H_A(t) + V(t)$, and can in principle have any kind of time-dependence. 
The process will thus in general involve the expenditure of work, as well as the exchange of heat between $S$ and $A$.
The interaction lasts for a time $\tau$, after which their joint state will be $\rho_{SA}(\tau) = U_\tau (\rho_S \otimes \rho_A) U_\tau^\dagger$, where $U_\tau = \mathcal{T} e^{-i \int_0^\tau dt~H(t)}$ and $\mathcal{T}$ is the time-ordering operator. 
The reduced state of the system will then be described by the quantum channel $\rho_S(\tau) = \Lambda[\rho_S]$, 
where $\Lambda[\bullet] = \tr_A U_\tau \big(\bullet \otimes \rho_A\big) U_\tau^\dagger$. If one is concerned with a closed system, $\Lambda$ entails the open system dynamics due to the presence of the measurement device and, if the system is already open, $\Lambda$ can include both dynamics and measurement backaction.

Our main interest is in the \emph{changes} undergone by some system observable $G_t$;
this could be, for instance, the local system energy $H_S(t)$.
The average change in $G$ is, of course,
\begin{equation}\label{ave_G}
    \langle \Delta G \rangle = \tr\Big\{ G_\tau \Lambda[\rho_S] - G_0 \rho_S\Big\}.
\end{equation}
But our interest is in going beyond the average, and account for the fluctuations.
We do this using the  two-point measurement  (TPM) protocol~\cite{Talkner2007}.
Let $G_0 = \sum_{g_0} \lambda_{g_0} |g_0\rangle\langle g_0|$ and $G_\tau = \sum_{g_\tau} \lambda_{g_\tau} |g_\tau\rangle \langle g_\tau|$ denote the eigendecompositions of $G_t$ at $t=0$ and $t=\tau$. 
For now, we assume that $[G_0, \rho_S] = 0$; the case where this does not hold is subtle, and is discussed below.
The TPM protocol consists in measuring the system in the eigenbases $\{|g_0\rangle\}$ and $\{|g_\tau\rangle\}$, respectively before and after the channel $\Lambda[\bullet]$. This leads to the distribution
\begin{equation}
    P(\Delta G) = \sum\limits_{g_0, g_\tau} \langle g_\tau| \Lambda\Big[|g_0\rangle\langle g_0| \Big] | g_\tau \rangle p(g_0) ~\delta\Big(\Delta G - (g_\tau - g_0) \Big).
\end{equation}
where $p(g_0) = \langle g_0 | \rho_S | g_0\rangle$. 
From this, higher order statistics can be readily computed. 
Of particular interest is the variance $\Var(\Delta G)= \langle \Delta G^2 \rangle - \langle \Delta G \rangle^2$, a quantity which has seen a surge of interest, e.g. in connection with the so-called thermodynamic uncertainty relations~\cite{Barato2015,Gingrich2016,Pietzonka2015, Timpanaro2019,Guarnieri2019,Hasegawa2020,Hasegawa2019a}.


\section{Predictors}  In this paper we assume that the distribution $P(\Delta G)$ is not accessible. Instead, all one has access to is a specific unravellings of the channel $\Lambda$: \begin{equation}\label{channel}
    \Lambda [\bullet] := \sum\limits_\gamma M_\gamma \bullet M_\gamma^\dagger,
    \qquad \sum_\gamma M_\gamma^\dagger M_\gamma = 1,
\end{equation}
where $M_\gamma = \langle \gamma | U_\tau | 0\rangle$ (we take $\rho_A = |0\rangle\langle 0|$ wlog~\cite{Nielsen}). The unravellings fix a POVM, $E_\gamma = M_\gamma^\dagger M_\gamma$, representing a set of measurements one has access to, whose outcomes occur with probability,
\begin{equation}\label{P_gamma}
    P(\gamma) = \tr\big( M_\gamma \rho_S M_\gamma^\dagger\big).
\end{equation}
Any function $\Delta \mathcal{G}(\gamma)$, of the stochastic outcomes $\gamma$, can now be viewed as a predictor of $\Delta G$, in the sense that it conveys some information about it.
Our goal is to determine which function $\Delta \mathcal{G}_{\rm opt}$ which yields the best possible prediction.

We quantify the quality of the prediction in terms of the mean-squared error (MSE)
\begin{equation}\label{MSE}
    {\rm MSE}\Big( \Delta \mathcal{G}\Big) = \sum\limits_\gamma \int d\Delta G~ \Big(\Delta \mathcal{G}(\gamma) - \Delta G\Big)^2 P(\Delta G, \gamma),
\end{equation}
where 
\begin{equation}\label{P_deltaG_gamma}
    P(\Delta G, \gamma) = \sum\limits_{g_0, g_\tau} |\langle g_\tau| M_\gamma|g_0\rangle|^2 p(g_0) ~\delta\Big(\Delta G - (g_\tau - g_0) \Big),
\end{equation}
is the joint distribution of $\Delta G$ and $\gamma$, which would have been obtained if both $S$ and $A$ had been measured. 
The main result of this letter is:
\\
\noindent
{\bf Theorem:} \emph{The predictor minimizing the MSE~\eqref{MSE} is given by 
\begin{equation}\label{pred_opt}
    \Delta \mathcal{G}_{\rm opt}(\gamma) = \frac{1}{P(\gamma)}  \left\langle M_\gamma^\dagger G_\tau M_\gamma - \frac{1}{2} \{M_\gamma^\dagger M_\gamma, G_0\}\right\rangle,
\end{equation}
where the average is over the system's initial state $\rho_S$.
This predictor always reproduces the true average~\eqref{ave_G}:
\begin{equation}\label{ave_G_opt}
    \sum\limits_\gamma P(\gamma) \Delta \mathcal{G}_{\rm opt}(\gamma) = \langle \Delta G \rangle.
\end{equation}
Moreover, it yields the MSE:
\begin{equation}\label{MSE_opt}
    {\rm MSE}(\Delta \mathcal{G}_{\rm opt}) = {\rm Var}(\Delta G) - {\rm Var}(\Delta \mathcal{G}_{\rm opt}),
\end{equation}
which is thus simply the difference between the  fluctuations of the true quantity and those of the predictor.
}

The proof of this result given in the Appendix~\ref{appA}.
The fact that the optimal predictor always correctly reproduces the correct average behavior is noteworthy. 
In addition, it yields an MSE which directly links to the variance of $\Delta G$.
In fact, Eq.~\eqref{MSE_opt}  implies that if $\Delta G$ does not fluctuate ($\Var(\Delta G) \equiv 0$), the same will also be true for $\Delta \mathcal{G}_{\rm opt}$ (since the MSE is strictly non-negative). 
When applied in a thermodynamic scenario (as will be done below), this result is consistent with the no-go theorem of Ref.~\cite{Perarnau-Llobet2017}.

\section{Energy fluctuations in a driven qubit}
To illustrate the idea, we consider a qubit system coupled to a qubit ancilla, with total Hamiltonian $H(t) = H_S(t) + \omega \sigma_z^A + g (\sigma_+^S \sigma_-^A + \sigma_-^S \sigma_+^A)$ and 
\begin{equation}\label{qubit_drive}
    H_S(t) = \omega \sigma_z^S + \alpha \sin(\Omega t) \sigma_x^S. 
\end{equation}
The system is thus driven by a time-dependent horizontal field $\alpha \sin(\Omega t) \sigma_x$. 
We look for estimates of how the system energy ($G_t \equiv H_S(t)$), changes from time 0 up to some generic time $\tau$. 

\begin{figure}
    \centering
    \includegraphics[width=0.5\textwidth]{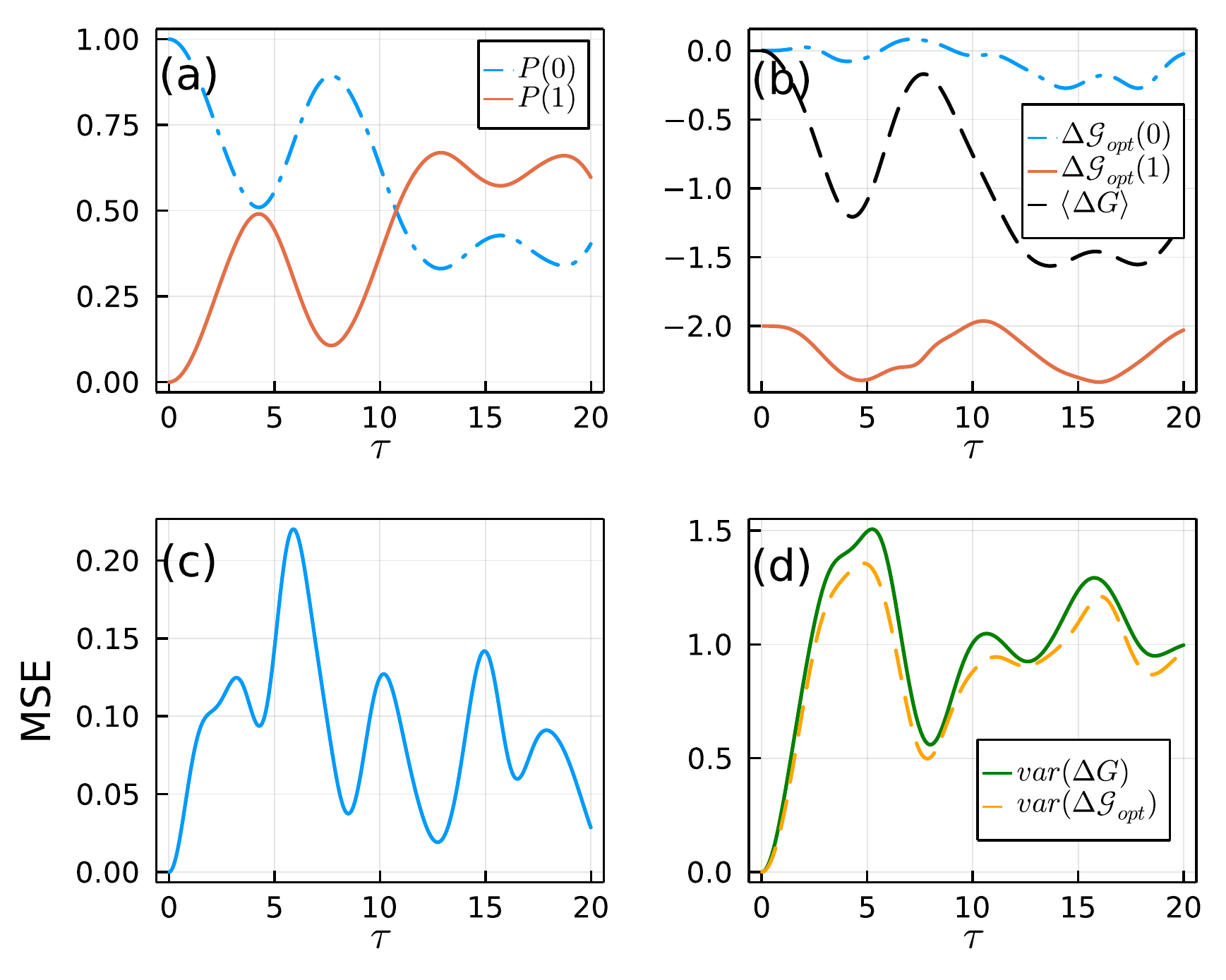}
    \caption{Driven qubit model, as a function of the net driving time $\tau$.
    (a) Probabilities $P(\gamma)$, Eq.~\eqref{P_gamma}; 
    (b) Optimal predictors $\Delta \mathcal{G}_{\rm opt}(\gamma)$, Eq.~\eqref{pred_opt}. 
    (c) Mean-squared error, Eq.~\eqref{MSE}.
    (d) True and predicted variances, $\Var(\Delta G)$ and $\Var(\Delta \mathcal{G}_{\rm opt})$. 
    Parameters: $\omega = \alpha = 1$, $g = \Omega =  0.3$, $s = 0.7$.    
    }
    \label{fig:qubit}
\end{figure}

The system starts in $\rho_S = s|0\rangle\langle 0| + (1-s)|1\rangle\langle 1|$, while the ancilla starts in $\rho_A = |0\rangle\langle 0|$. 
Moreover, we measure the ancillas in the computational basis, leading to two possible Kraus operators $M_0 = \langle 0| U_\tau| 0\rangle$ and $M_1 = \langle 1 | U_\tau |0 \rangle$, where $U_\tau$ is obtained numerically.
The results are shown in Fig.~\ref{fig:qubit}:
The probabilities $P(\gamma)$ in Eq.~\eqref{P_gamma} evolve as shown in Fig.~\ref{fig:qubit}(a). 
And for each such $\gamma$, Eq.~\eqref{pred_opt} predicts the results shown in Fig.~\ref{fig:qubit}(b). 
We also plot in black-dashed lines the true average $\langle \Delta G\rangle$ [c.f. Eq.~\eqref{ave_G_opt}].

We can estimate the quality of our predictions by plotting the MSE~\eqref{MSE} (Fig.~\ref{fig:qubit}(c)). 
The predictions are better at some values of $\tau$ and worse at others. 
This can also be inferred by comparing the actual variance of $\Delta G$ with the predicted one $\Var(\Delta \mathcal{G}_{\rm opt})$, as shown in Fig.~\ref{fig:qubit}(d). 
This result makes it particularly clear that very good predictions are possible, despite the complexity of the problem. 
Of course, this will depend on the model and the parameters.

\section{Interpretation, particular cases and generalizations} The optimal predictor is a function  $f(\gamma)$, which provides the best possible guess for the changes in the system observable, from $G_0$ to $G_\tau$, given only the specific unravelling in the ancilla. 
Computing it thus requires a \emph{model}.
In classical statistics, this is usually associated with a transition probability $P(y|x)$. 
In our case, this is determined  by the the Kraus operators $\{M_\gamma\}$, as well as the system's initial state. 
In the way the problem has been  formulated, the operator $G$ can represent any system property. Hence, the results extend well beyond thermodynamics.
The unique thermodynamic flavor appears in the fact that we are not estimating the value of $G$ itself, but rather the \emph{change} in $G$ due to a certain process. 
In fact, if we artificially adjust $G_0 \equiv 0$, Eq.~\eqref{pred_opt} reduces to  $\Delta \mathcal{G}_{\rm opt}(\gamma) = \tr(G_\tau \rho_{\tau|\gamma})$, where $\rho_{\tau|\gamma} = M_\gamma \rho M_\gamma^\dagger/P(\gamma)$ is the conditional state, given $\gamma$. 
In this case the predictor is therefore only estimating the value of $G_\tau$ itself. 
Similarly, if $G_\tau=0$ Eq.~\eqref{pred_opt} will provide a retrodiction~\cite{Barnett2021}, of the initial value $G_0$ before the open dynamics.
We also note that the choice of unravelling in the ancilla is arbitrary and model dependent. Hence, some unravellings will be more precise than others. 
This can be quantified using the MSE~\eqref{MSE_opt}.

Eq.~\eqref{pred_opt} can be readily extended to more complex types of unravellings. 
For instance, continuous measurements will be characterized by a string of indices $\gamma = (\gamma_1, \ldots, \gamma_N)$. 
Eq.~\eqref{pred_opt} still applies. But now the Kraus operators will have the form $M_\gamma = M_{\gamma_N} \ldots M_{\gamma_1}$. 
Predictions can also be made when only part of the outcomes are known. For instance, if there are two outcome $\gamma_1, \gamma_2$, but only the latter is known, then the optimal predictor will be simply $\Delta \mathcal{G}(\gamma_2) = \sum_{\gamma_1} P(\gamma_1) \Delta\mathcal{G}_{\rm opt}(\gamma_1,\gamma_2)$.

\section{Thermodynamic applications }
To make a connection with thermodynamics, we consider the problem of determining the work associated to an avoided crossing protocol~\cite{Pekola2013}.
We consider a qubit with $\rho_S = s|0\rangle\langle 0| + (1-s)|1\rangle\langle 1|$ and time-independent Hamiltonian $H_S = \omega \sigma_+ \sigma_-$. 
Initially the qubit is isolated and undergoes a unitary pulse described by $U_{\rm w} = \sigma_x$.
The work associated to this pulse can be either $\pm \omega$. 
In order to estimate this, 
we then couple the system to an ancilla, \emph{after} the protocol. For simplicity, we take the ancilla at zero-temperature, $\rho_A = |0\rangle\langle 0|$; and take $H_A = \omega |1\rangle\langle 1|$.
Moreover, the interaction is assumed to be a swap unitary, which will hence transfer any excitations from $S$ to $A$. 
Overall, this will thus be a two-step process, with a net unitary $U = U_{\rm swap}(U_w \otimes I_A)$.

After the process, we measure the ancilla in the computational basis, which leads to two Kraus operators $M_\gamma = \langle \gamma | U_{\rm swap} |0\rangle U_{\rm w}$, with $\gamma = 0,1$. 
We use the predictor~\eqref{pred_opt}, with $G_0 = G_\tau = H_S$. 
To obtain the actual work performed in the system, we must also include the heat transferred to the bath.
In more complicated scenarios, such as strong coupling, one might also need to include the system-ancilla interaction. Our formalism currently cannot account for this. 
In the present case, however, the swap incurs no additional energy cost since the system and ancilla are resonant. Hence, the work will simply be given by $W = Q + \Delta H_S$. Consequently, the optimal predictor will be 
\begin{equation}\label{pred_work}
    \mathcal{W}_{\rm opt}(\gamma) = Q(\gamma) + \Delta \mathcal{G}_{\rm opt}(\gamma),
\end{equation}
where, in this case, $Q(1) = \omega$ and $Q(0) = 0$.
There are thus two possibilities (Table~\ref{tab:minimal_qubit_model}). If $M_1$ is detected, the ancilla must have absorbed an excitation, which  means that the work protocol must have excited the system from $|0\rangle \to |1\rangle$.
The work associated to this is thus $+\omega$. 
Conversely, $M_0$ means that no excitation was detected in the ancilla, so that the state of the system before the swap must have been $|0\rangle$. 
Hence, the work performed must have been $-\omega$.

\begin{table}
    \centering
    \caption{Trajectories, probabilities and predicted work [Eq.~\eqref{pred_work}] for the avoided crossing model. Here $\Delta \mathcal{G}_{\rm opt}$ is the predictor for the system energy, while $\mathcal{W}_{\rm opt}$ refers to the work, which also includes the heat, as in Eq.~\eqref{pred_work}.
    }
    \begin{tabular}{l|c|c|c|c}
    \\[0.2cm]
      $M_\gamma$ \;\;
      &  \;\;$P[\gamma]$ \;\;
      & \;\;$\Delta \mathcal{G}_{\rm opt}(\gamma)$\;\; 
      & \;\;$\mathcal{W}_{\rm opt}(\gamma)$\;\; 
      & \;\; $\mathcal{W}_{\rm opt}(\gamma)$ (coherent) \;\; 
      \\[0.2cm]\hline
      $M_0$ 
      & $s$
      & $-\omega$ 
      & $-\omega$
      & $-\omega + \frac{\omega}{2} \frac{\sin^2\theta}{1+ (2s-1)\cos\theta}$ 
      \\[0.2cm]
      $M_1$ 
      &$1-s$ 
      & $0$ 
      & $\omega$
      &$\omega - \frac{\omega}{2} \frac{\sin^2\theta}{1- (2s-1)\cos\theta}$ 
    \end{tabular}
    \label{tab:minimal_qubit_model}
\end{table}

\section{Initially coherent systems}
The optimal predictor~\eqref{pred_opt} assumes that $[G_0, \rho_S] = 0$.
When this is not the case, the  problem becomes more delicate~\cite{Perarnau-Llobet2017}, and  several approaches have been put forward for handling it~\cite{Allahverdyan2005a,Engel2007,Allahverdyan2014,Hofer2016,Solinas2016,Hofer2017,Miller2017,Solinas2017,YungerHalpern2018a,Levy2019,Park2017,Micadei2020,Guerardini2020}.
Within our framework, the problem remains essentially unaltered, in the sense that the quality of any prediction is still given by the MSE~\eqref{MSE}.
The subtle part is in how to define the joint distribution $P(\Delta G, \gamma)$, as this would generally be susceptible to the backaction from the first measurement. 
A way of constructing $P(\Delta G, \gamma)$, which does not suffer from this problem is through the concept of quantum Bayesian networks (QBNs)~\cite{Park2017,Micadei2020}.
Let $\rho_S = \sum_\alpha p_\alpha |\psi_\alpha\rangle\langle \psi_\alpha|$, with the bases $\{|\psi_\alpha\rangle\}$ and $\{|g_0\rangle\}$ being generally incompatible. 
We then consider~\cite{Park2017,Micadei2020}
\begin{equation}\label{P_gamma_aug}
    P(\Delta G, \gamma) = \sum\limits_{g_0, g_\tau, \alpha} |\langle g_\tau |M_\gamma | \psi_\alpha\rangle|^2 p_{g_0|\alpha} p_\alpha \delta\Big(\Delta G - (g_\tau - g_0)\Big),
\end{equation}
where $p_{g_0|\alpha} = |\langle g_0 |\psi_\alpha\rangle|^2$ is the conditional probability of observing $|g_0\rangle$ given $|\psi_\alpha\rangle$. 
The QBN (i) is always non-negative, (ii) reproduces the correct average~\eqref{ave_G} and (iii) reduces to the TPM~\eqref{P_deltaG_gamma} when $[G_0,\rho_S] = 0$. 
From an operational perspective, it was also recently shown that QBNs can be directly accessed in an experiment, provided one uses two identical copies of the system, together with measurement post-processing~\cite{Micadei2021}. 

Plugging~\eqref{P_gamma_aug} in~\eqref{MSE} and repeating the same procedure in Appendix~\ref{appA} yields the optimal predictor 
\begin{equation}\label{pred_opt_coh}
    \Delta \mathcal{G}_{\rm opt}(\gamma) = \frac{1}{P(\gamma)}  \left\langle M_\gamma^\dagger G_\tau M_\gamma - \frac{1}{2} \{M_\gamma^\dagger M_\gamma, \mathbb{D}(G_0)\}\right\rangle,
\end{equation}
where $\mathbb{D}(\bullet) = \sum_\alpha |\psi_\alpha\rangle\langle \psi_\alpha| \bullet |\psi_\alpha\rangle\langle \psi_\alpha|$ is the full dephasing operator in the basis of $\rho_S$.
Compared to~\eqref{pred_opt}, the only difference is that $G_0$ is now replaced by $\mathbb{D}(G_0)$.
This therefore clearly reduces to~\eqref{pred_opt} when $[G_0, \rho_S] = 0$. 
Eq.~\eqref{pred_opt_coh} continues to yield the correct average, as in Eq.~\eqref{ave_G_opt}. 
And the MSE is still given by Eq.~\eqref{MSE_opt}.

\begin{figure}
    \centering
    \includegraphics[width=0.45\textwidth]{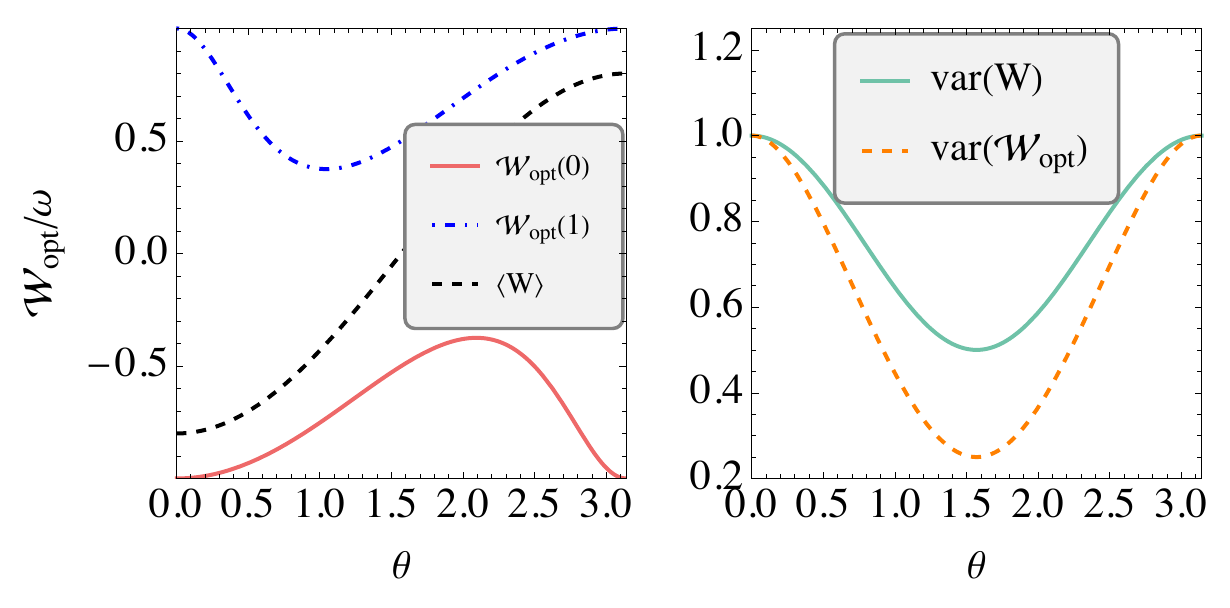}
    \caption{(a) Work predictions in the avoided crossing model (last column of Table~\ref{tab:minimal_qubit_model}), as a function of the system initial coherence angle $\theta$, with $s = 0.9$. (b) True and predicted variances. Their difference yields the MSE~\eqref{MSE_opt}, which quantifies the quality of the prediction.}
    \label{fig:avoided_crossing_pred}
\end{figure}

As an application, we revisit the avoided crossing model in Table~\ref{tab:minimal_qubit_model} and  include the effects of initial coherence, by assuming that the system is prepared in $\rho_S^{\rm coh} = e^{-i \theta \sigma_y/2} \rho_S e^{i \theta \sigma_y/2}$, where $\rho_S = s|0\rangle\langle 0| + (1-s)|1\rangle\langle 1|$.
The results are shown in the last column of Table~\ref{tab:minimal_qubit_model}, as well as in  
Fig.~\ref{fig:avoided_crossing_pred}(a). 
In the latter, we also plot the average work $\langle W \rangle$ for comparison. 
To assess the quality of the prediction, we plot in Fig.~\ref{fig:avoided_crossing_pred}(b) the true and predicted variances, $\Var(W)$ and $\Var(\mathcal{W}_{\rm opt})$. 
They are equal in the incoherent case and grow with increasing $\theta$, being maximal when $\theta = \pi/2$.
The presence of quantum coherences therefore generally degrades the quality of the prediction, which is intuitive, although this may not be true for other models.

\section{Discussion}
We have put forth a  framework for finding the optimal function to predict changes of a system property indirectly, solely from measurements in an auxiliary system.
The method is applicable to both coherent and incoherent systems, and is summarized by an explicit expression for the predictor in terms solely of the Kraus operators determining the open system unravelling.
The predictor can always capture the correct average. And the quality of the estimation is directly related to how well it captures the variance.
These results are timely, due to the growing progress in the manipulation of coherent quantum devices and their potential thermodynamic applications. 
To illustrate that, we analyzed the problem of estimating energy fluctuations in driven qubit systems, and in the reconstruction of work in an avoided crossing model.

\begin{acknowledgments}
The authors thank E. Lutz for the feedback, and K. Micadei, F. Rodrigues, L. Jenkins and T. A. Pinto Silva for fruitful discussions. 
GTL acknowledges the support from the S\~ao Paulo Funding Agency FAPESP, under grants 2018/12813-0, 2017/50304-7 and 2017/07973-5.
MJ acknowledges the financial support from 
the Brazilian funding agency CAPES.
\end{acknowledgments}

\appendix
%
%
\section{Proof of the main theorem}\label{appA}
%
%

We prove here the theorem in Eq.~(7) of the main text; namely, that the predictor $\Delta \mathcal{G}_{\rm opt}(\gamma)$ which minimizes the mean-squared error~\eqref{MSE} is that given by Eq.~\eqref{pred_opt}. 
Define 
\begin{equation}\label{SupMat_f_def}
    f(\gamma) = \frac{1}{P(\gamma)} \int d\Delta G~(\Delta G) P(\Delta G, \gamma) 
\end{equation}
We consider the MSE in Eq.~\eqref{MSE} for a generic predictor $\Delta \mathcal{G}$. 
Adding and subtracting $f(\gamma)$ leads to 
\begin{IEEEeqnarray}{rCl}
{\rm MSE}(\Delta \mathcal{G}) &=& 
\sum\limits_\gamma \int d\Delta G~\Big( \Delta G - f(\gamma)\Big)^2 P(\Delta G, \gamma)
\\[0.2cm]
\nonumber 
&&\sum\limits_\gamma \int d\Delta G~\Big( \Delta \mathcal{G}(\gamma) - f(\gamma)\Big)^2 P(\Delta G, \gamma)
\\[0.2cm]
\nonumber 
&&
\sum\limits_\gamma \int d\Delta G~\Big( \Delta \mathcal{G}(\gamma) - f(\gamma)\Big)\Big( f(\gamma) - \Delta G\Big) P(\Delta G, \gamma)
\end{IEEEeqnarray}
Due to~\eqref{SupMat_f_def}, however, the last term vanishes. Moreover, in the second term we can marginalize over $\Delta G$ and write the result as an average over the original distribution $P(\gamma)$:
\begin{IEEEeqnarray}{rCl}\label{SM_MSE_gen_2}
{\rm MSE}(\Delta \mathcal{G})  &=& \sum\limits_\gamma \int d\Delta G~\Big( \Delta G - f(\gamma)\Big)^2 P(\Delta G, \gamma)
\\[0.2cm]
\nonumber
&&+ \sum\limits_\gamma P(\gamma) \Big( \Delta \mathcal{G}(\gamma) - f(\gamma)\Big)^2.
\end{IEEEeqnarray}
Since the last term is always non-negative, comparing this with Eq.~\eqref{MSE} leads to
\begin{equation}
    {\rm MSE}(\Delta \mathcal{G}) \geqslant {\rm MSE}(f),
\end{equation}
for any other predictor $\Delta \mathcal{G}$.
Hence, the optimal predictor is exactly $\Delta \mathcal{G}_{\rm opt}(\gamma) \equiv f(\gamma)$. 

For the optimal predictor the MSE~\eqref{SM_MSE_gen_2} reduces to 
\begin{equation}
    {\rm MSE}(\Delta \mathcal{G})  = \sum\limits_\gamma \int d\Delta G~\Big( \Delta G - f(\gamma)\Big)^2 P(\Delta G, \gamma).
\end{equation}
Expanding and using Eq.~\eqref{SupMat_f_def} leads to Eq.~\eqref{MSE_opt} of the main text. 

Next, we obtain the explicit form in Eq.~\eqref{pred_opt}, in which the optimal predictor is cast solely in terms of the Kraus operators. 
Inserting Eq.~\eqref{P_deltaG_gamma} into Eq.~\eqref{SupMat_f_def}
leads to 
\begin{equation}
    \Delta \mathcal{G}_{\rm opt}(\gamma) = \frac{1}{P(\gamma)}\sum\limits_{g_0,g_\tau} |\langle g_\tau| M_\gamma|g_0\rangle|^2 p(g_0) (g_\tau-g_0), 
\end{equation}
which can also be written as 
\begin{equation}
    \Delta \mathcal{G}_{\rm opt}(\gamma) = \frac{1}{P(\gamma)}\sum\limits_{g_\tau}
    g_\tau \langle g_\tau |M_\gamma \rho_S M_\gamma^\dagger | g_\tau\rangle  - \langle g_\tau| M_\gamma \rho_S G_0 M_\gamma^\dagger |g_\tau\rangle.
\end{equation}
Writing the remaining sum over $g_\tau$ as a trace yields precisely to Eq.~\eqref{pred_opt}. 
The coherent case, where $P(\Delta G, \gamma)$ is given by Eq.~\eqref{P_gamma_aug}, is treated similarly.

\bibliography{references}




\end{document}